\documentclass[11pt,a4paper]{article} 
\pdfoutput=1
\usepackage{jheppub,hyperref,mdwlist}
\usepackage{amsmath}
\usepackage{graphicx}
\usepackage{subfigure}
\usepackage{amssymb}
\usepackage{xcolor}
\usepackage{multirow}
\usepackage{cancel}
\usepackage{color}
\usepackage{listings}
\usepackage{xcolor}
\usepackage{float}
\usepackage{slashed}
\usepackage{mathtools}
\usepackage{tabularx}

\usepackage{tikz}
\usetikzlibrary{arrows,shapes}
\usetikzlibrary{trees}
\usetikzlibrary{matrix,arrows} 				
\usetikzlibrary{positioning}				
\usetikzlibrary{calc,through}				
\usetikzlibrary{decorations.pathreplacing}  
\usepackage{pgffor}							

\usetikzlibrary{decorations.pathmorphing}	
\usetikzlibrary{decorations.markings}
\tikzset{
    vector/.style={decorate, decoration={snake}, draw},
	provector/.style={decorate, decoration={snake,amplitude=2.5pt}, draw},
	antivector/.style={decorate, decoration={snake,amplitude=-2.5pt}, draw},
    fermion/.style={draw=black, postaction={decorate},
        decoration={markings,mark=at position .55 with {\arrow[draw=black]{>}}}},
    fermionbar/.style={draw=black, postaction={decorate},
        decoration={markings,mark=at position .55 with {\arrow[draw=black]{<}}}},
    fermionnoarrow/.style={draw=black},
    gluon/.style={decorate, draw=black,
        decoration={coil,amplitude=4pt, segment length=5pt}},
    scalar/.style={dashed,draw=black, postaction={decorate},
        decoration={markings,mark=at position .55 with {\arrow[draw=black]{>}}}},
    scalarbar/.style={dashed,draw=black, postaction={decorate},
        decoration={markings,mark=at position .55 with {\arrow[draw=black]{<}}}},
    scalarnoarrow/.style={dashed,draw=black},
    electron/.style={draw=black, postaction={decorate},
        decoration={markings,mark=at position .55 with {\arrow[draw=black]{>}}}},
	bigvector/.style={decorate, decoration={snake,amplitude=4pt}, draw},
}

\tikzstyle{block} = [draw, rectangle, 
    minimum height=3em, minimum width=6em]

\usepackage{amsmath}
\usepackage{wrapfig}
\usepackage{cleveref}

\newcommand{\be}{\begin{equation}}
\newcommand{\ee}{\end{equation}}
\newcommand{\beq}{\begin{equation}}
\newcommand{\eeq}{\end{equation}}
\newcommand{\bea}{\begin{eqnarray}}
\newcommand{\eea}{\end{eqnarray}}
\newcommand{\besp}{\begin{equation}\begin{split}}
\newcommand{\eesp}{\end{split}\end{equation}}

\newcommand{\Eq}[1]{Eq.~(\ref{#1})}

\newcommand{\Dfbd}{\mathord{\buildrel{\lower3pt\hbox{$\scriptscriptstyle\leftrightarrow$}}\over {D}_{\mu}}}
\newcommand{\ave}[1]{\left\langle #1\right\rangle}

\def\mE{\mathcal{E}}
\def\mF{\mathcal{F}}

\def\mL{\mathcal{L}}

\def\mO{\mathcal{O}}

\def\mQ{\mathcal{Q}}

\def\mS{\mathcal{S}}

\def\0{\textbf{0}}
\def\1{\textbf{1}}
\def\2{\textbf{2}}
\def\3{\textbf{3}}
\def\4{\textbf{4}}
\def\5{\textbf{5}}
\def\6{\textbf{6}}
\def\7{\textbf{7}}
\def\8{\textbf{8}}
\def\9{\textbf{9}}

\def\d{\text{d}}

\def\x{\textbf{x}}
\def\p{\textbf{p}}

\begin{document}

\title{Revisiting the fermion-field nontopological solitons}

\author[a]{Ke-Pan Xie}
\affiliation[a]{School of Physics, Beihang University, Beijing 100191, China}

\emailAdd{kpxie@buaa.edu.cn}

\abstract{Nontopological fermionic solitons exist across a diverse range of particle physics models and have rich cosmological implications. This study establishes a general framework for calculating fermionic soliton profiles under arbitrary scalar potentials, utilizing relativistic mean field theory to accurately depict the interaction between the fermion condensate and the background scalar field. Within this framework, the conventional ``fermion bound states'' are revealed as a subset of fermionic solitons. In addition, we demonstrate how the analytical formulae in previous studies are derived as special cases of our algorithm, discussing the validity of such approximations. Furthermore, we explore the phenomenology of fermionic solitons, highlighting new formation mechanisms and evolution paths, and reconsidering the possibility of collapse into primordial black holes.}

\maketitle
\flushbottom

\section{Introduction}

Numerous particle physics models predict extended and localized states with mass varying from GeV to galactic level, dubbed solitons. Unlike the topological solitons such as cosmic strings or domain walls, nontopological solitons achieve stability not by vacuum topology, but by carrying conserved N\"{o}ther charges~\cite{Nugaev:2019vru,Lee:1991ax}. A prominent example is the Q-ball composed of scalar field with nonlinear interactions~\cite{Rosen:1968mfz,Friedberg:1976me,Coleman:1985ki}. The study on nontopological solitons involving fermion-field traces back to the work of T. D. Lee (with his collaborators) and E. Witten in the 1970s  and 1980s~\cite{Friedberg:1976eg,Friedberg:1977xf,Lee:1986tr,Witten:1984rs}. Further studies have unveiled that fermionic solitons can form abundantly in the Universe with profound cosmological implications, such as being dark matter and/or generating the matter-antimatter asymmetry~\cite{Bai:2018vik,Bai:2018dxf,Hong:2020est,Marfatia:2021twj,Gross:2021qgx,Macpherson:1994wf,Oaknin:2003uv,Atreya:2014sca,Liang:2016tqc,Ge:2017idw,Ge:2019voa,Zhitnitsky:2021iwg}, and sourcing primordial black holes~\cite{Kawana:2021tde,Huang:2022him,Kawana:2022lba,Xie:2023cwi,Marfatia:2021hcp,Marfatia:2022jiz,Lu:2022jnp,Tseng:2022jta,Acuna:2023bkm,Chen:2023oew,Kim:2023ixo,Borah:2024lml}.

\begin{figure}
\begin{center}
\includegraphics[width=0.99\textwidth]{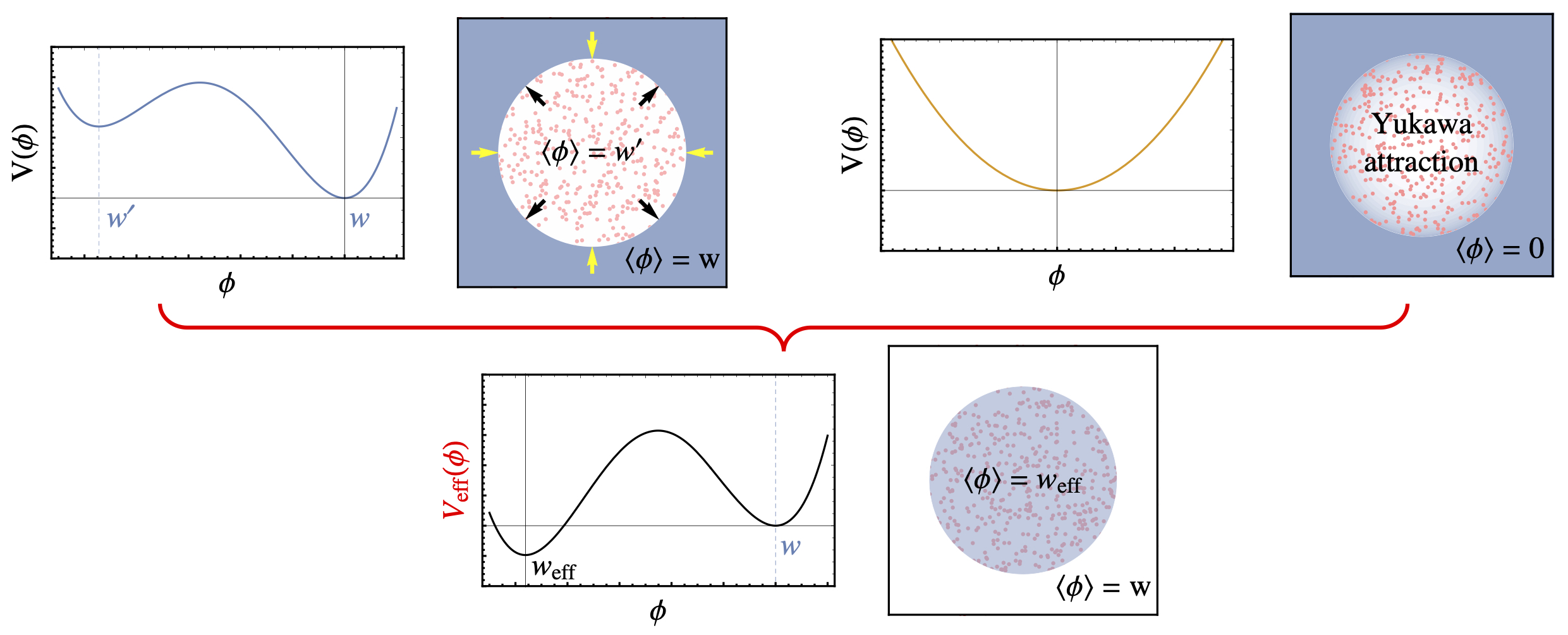}
\caption{Illustration of physical pictures. {\bf Top-left}: the conventional fermionic soliton, where fermions are trapped in the false vacuum $w'$ of $V(\phi)$, and the degeneracy pressure balances the vacuum pressure. {\bf Top-right}: the conventional fermion bound state, where $V(\phi)\sim m'^2\phi^2/2$ features only one vacuum, and fermions are bound by the Yukawa attraction mediated by $\phi$. {\bf Bottom}: the new perspective in this work, where both the conventional ``solitons'' and ``bound states'' are described by an effective potential $V_{\rm eff}(\phi)$ incorporating the fermionic $\ave{\bar\chi\chi}$ contribution. The soliton corresponds to the \underline{true} vacuum $w_{\rm eff}$ of $V_{\rm eff}(\phi)$.}
\label{fig:sketch}
\end{center}
\end{figure}

The conventional description of fermionic soliton is a tale of the balance between the degeneracy and vacuum pressures, as sketched in the top-left panel of Fig.~\ref{fig:sketch}. The basic setup involves a real scalar $\phi$ and a fermion $\chi$ interacting via Yukawa coupling, and a scalar potential $V(\phi)$ featuring multiple vacua (minima) separated by barrier(s).\footnote{Even in the strong dynamics models without elementary scalars, effective scalar degrees of freedom emerge from the quark condensate $\ave{\bar qq}$~\cite{Bai:2018dxf} or Polyakov loop constructed by gauge fields~\cite{Pasechnik:2023hwv}.} The global minimum is the true vacuum. The energetically unstable false vacuum region in the space shrinks due to inward vacuum pressure. However, if identical fermions are trapped within a false vacuum region, contraction stops when the outward degeneracy pressure balances the vacuum pressure, resulting in soliton formation. The constituent fermions are trapped in the soliton due to their increased mass in the true vacuum, as they lack sufficient kinetic energy to overcome the mass gap and escape.

Previous research exhibits two features. First, polynomial potentials serve as the primary benchmark for study. Ref.~\cite{Friedberg:1976eg} proves the existence theorem for fermionic solitons under $V(\phi)=a\phi^2/2+b\phi^3/3!+c\phi^4/4!$ with $a>0$, $c\geqslant0$, and $b^2\leqslant 3ac$, which ensures the co-existence of two vacua. Investigations into various polynomial potentials have been conducted~\cite{Lee:1986tr,Koeppel:1985tt,DelGrosso:2023trq,DelGrosso:2023dmv,DelGrosso:2024wmy}. However, many well-motivated scenarios involve non-polynomial potentials, such as logarithmic potential in classically conformal theories~\cite{Iso:2009ss,Iso:2009nw,Chun:2013soa}, and non-analytical thermal corrections in finite-temperature field theories~\cite{Dolan:1973qd,Quiros:1999jp}. Therefore, techniques for solving fermionic solitons under a general $V(\phi)$ are imperative. Second, most phenomenological studies~\cite{Bai:2018vik,Bai:2018dxf,Hong:2020est,Marfatia:2021twj,Gross:2021qgx,Macpherson:1994wf,Oaknin:2003uv,Atreya:2014sca,Liang:2016tqc,Ge:2017idw,Ge:2019voa,Zhitnitsky:2021iwg,Kawana:2021tde,Huang:2022him,Kawana:2022lba,Xie:2023cwi,Marfatia:2021hcp,Marfatia:2022jiz,Lu:2022jnp,Tseng:2022jta,Acuna:2023bkm,Chen:2023oew,Kim:2023ixo,Borah:2024lml} assume a uniform spatial distribution of $\chi$ particles within the soliton, resulting in profiles as analytical functions of the soliton charge and radius. However, this approximation overlooks the fact that the dense fermions inside the soliton modifies $V(\phi)$, subsequently affecting fermion mass and leading to a nonuniform distribution of $\chi$. A more precise treatment including those effects is necessary.

This work revisits the fermionic soliton scenario and improves the calculation of the soliton profile. Moving beyond the polynomial ansatz, I consider a general potential $V(\phi)$ and take into account the influences between the $\ave{\bar\chi\chi}$ condensate and the $\phi$ field. Employing relativistic mean field theory in Minkowski spacetime, I demonstrate that the formation of fermionic solitons necessitates the multi-vacuum structure not of $V(\phi)$ but of $V_{\rm eff}(\phi)$, an effective potential including the fermion contribution. The soliton lives in the \underline{true} vacuum of $V_{\rm eff}(\phi)$ rather than the \underline{false} vacuum of $V(\phi)$. An implementable algorithm is introduced for evaluating the soliton profile, and the analytical formulae observed in previous studies are reproduced as special cases. The validity conditions of analytical approximations are provided.

The new algorithm not only enhances methodology but also provides new insights into the concept of fermionic solitons by encompassing the conventional {\it fermion bound state} as a subset. Traditionally, fermionic solitons and bound states are perceived as distinct entities: although both form via Yukawa interactions, the former needs $V(\phi)$ to have a multi-vacuum structure, while the latter features a single-vacuum $V(\phi)\sim m'^2\phi^2/2$ requiring $m'$ to be sufficiently small. This allows $\phi$ mediate a long-range attractive force between the fermions, leading to the formation of bound states akin to atoms or nuclei~\cite{Stephenson:1996qj,Wise:2014jva,Wise:2014ola,Gresham:2017zqi,Gresham:2018anj,Smirnov:2022sfo}, as sketched in the top-right panel of Fig.~\ref{fig:sketch}. However, the framework established in this study reveals that the treatment of those two types of objects is exactly the same. In particular, it is $V_{\rm eff}(\phi)$, not $V(\phi)$, that is crucial for soliton formation. Even if $V(\phi)$ is a single-vacuum potential, the fermion source can deform the potential, resulting in a multi-vacuum $V_{\rm eff}(\phi)$. Therefore, there is no intrinsic difference between solitons and bound states; both of them can be viewed as fermions living in the \underline{true} vacuum of $V_{\rm eff}(\phi)$. This is sketched in the bottom panel of Fig.~\ref{fig:sketch}.

Based on the new understanding of fermionic solitons, this work offers a comprehensive discussion of their phenomenology. This includes elucidating various formation mechanisms, exploring potential evolution paths and cosmological implications, alongside a discussion on detectable astrophysical and particle experiment signals. Notably, contrary to early studies, it is observed that solitons are unlikely to collapse into primordial black holes via the internal Yukawa force, attributed to the limitations of the Yukawa attraction approximation.

This paper is organized as follows. The equations of motion (EoMs) for soliton formation are derived in section~\ref{sec:framework}. Then section~\ref{sec:solution} details the resolution of the EoMs, outlining the general algorithm and discussing analytical limits. Two numerical examples are offered in section~\ref{sec:numerical}. Section~\ref{sec:pheno} explores phenomenological aspects. Finally, the conclusion and outlook are given in section~\ref{sec:conclusion}. Before proceeding, we briefly address the terminology used in this article. Various terms exist in the literature for such balls composed of confined fermions and nontrivial scalar field configuration, including ``fermion-field nontopological solitons'' or simply ``fermion solitons''~\cite{Friedberg:1976eg,Friedberg:1977xf,Lee:1986tr}, or ``quark nugget''~\cite{Witten:1984rs} and ``neutrino-ball''~\cite{Holdom:1987ep} according to the fermion constituents. To streamline our discussion, we adopt the term ``Fermi-ball'' following Ref.~\cite{Hong:2020est}.

\section{The equations of motion}\label{sec:framework}

The prototype Lagrangian reads
\be\label{Yuk_bound_L}
\mL=\frac12\partial_\mu\phi\partial^\mu\phi-V(\phi)+\bar\chi i\slashed{\partial}\chi-y\phi\bar\chi\chi,
\ee
where $\chi$ is the fermion, and $\phi$ is the real scalar. The bare Dirac mass of $\chi$ has been absorbed into the definition of $V(\phi)$, i.e. $-M_f\bar\chi\chi$ can be formally eliminated via the field shift $\phi\to(\phi-M_f/y)$. The scalar potential is not specified and it may have one or more minima. For simplicity, we assume there are at most two local minima; the global minimum (true vacuum) is located at $w>0$ with $V(w)=0$, at which $\chi$ acquires a mass $M=yw$. If there exists a second local minimum (false vacuum), it is located at $w'\in[0,w)$ and satisfies $V(w')\geqslant0$. This implies $\d V/\d\phi|_{\phi=0}\leqslant0$. The findings presented in this article are easily generalized to scalar potentials with three or more minima.

The EoMs from the Lagrangian \Eq{Yuk_bound_L} are obtained by variation principle,
\be\label{bare_EoM}
\partial_\mu\partial^\mu\phi+\frac{\partial V}{\partial \phi}=-y\bar\chi\chi,\quad i\slashed{\partial}\chi=y\phi\chi,
\ee
where the first one describes the fermions are the source of the scalar field, while the second one describes the fermion motion is affected by the Yukawa force. Since the Lagrangian is invariant under the $U(1)$ transformation $\chi\to e^{i\theta}\chi$, a conserved current $J^\mu=\bar\chi\gamma^\mu\chi$ exists in the motion, and the conserved charge is
\be
Q=\int\d^3xJ^0=\int\d^3x\chi^\dagger\chi,
\ee
which is the number difference between $\chi$ and its antiparticle $\bar\chi$.

The EoMs in \Eq{bare_EoM} are for the field operators in the Heisenberg picture and can be simplified in the framework of relativistic mean field theory. Assuming a large field occupancy, the scalar can be treated as a static classical background field $\phi(\x)$, and the fermion source in the first EoM can be replaced by the ensemble average $\ave{\bar\chi\chi}$. Additionally, assuming the scalar field varies slowly, the second EoM introduces a space-dependent mass term $m(\x)=y\phi(\x)$ for the fermions. A Fermi-ball should consist of only $\chi$ or $\bar\chi$ particles, otherwise the $\chi\bar\chi$ annihilations will destabilize the object. Without loss of generality, we assume the soliton contains only $\chi$'s, and hence the soliton charge $Q$ equals to the number of the constituent fermions. The phase space distribution follows the parametrized Fermi-Dirac statistics
\be
f(\x,\p)=\frac{1}{e^{(\epsilon-\epsilon_F)/T}+1},
\ee
where $T$ is the temperature, $\epsilon_F(\x)$ is a space-dependent parameter, and the single particle energy is $\epsilon(\x,\p)=\sqrt{\p^2+m^2(\x)}$. 

Thermodynamic observables of the system are expressed as functionals of $\epsilon_F(\x)$ and $\phi(\x)$. The charge is
\be
\mQ[\phi,\epsilon_F]=g_{\rm dof}\int\d^3x\int\frac{\d^3p}{(2\pi)^3}f(\x,\p),
\ee
where $g_{\rm dof}$ is the spin and internal space degeneracy of $\chi$. The total energy (also the mass of a soliton at rest) is
\be
\mE[\phi,\epsilon_F]=\int\d^3x\left[\frac12(\nabla\phi)^2+V(\phi)+g_{\rm dof}\int\frac{\d^3p}{(2\pi)^3}\epsilon f\right],
\ee
which is the summation of the scalar and the fermion components. Explicitly, the entropy is solely contributed by the $\chi$ particles,
\be
\mS[\phi,\epsilon_F]=\frac{g_{\rm dof}}{T}\int\d^3x\int\frac{\d^3p}{(2\pi)^3}(\epsilon-\epsilon_F)f+g_{\rm dof}\int\d^3x\int\frac{\d^3p}{(2\pi)^3}\log\left(1+e^{-(\epsilon-\epsilon_F)/T}\right),
\ee
because the classical field $\phi(\x)$ has no entropy, but it affects $\mS[\phi,\epsilon_F]$ implicitly via the influence on $\chi$ mass.

According to the ensemble theory, for a given charge $\mQ[\phi,\epsilon_F]=Q$, the functions $\epsilon_F(\x)$ and $\phi(\x)$ should minimize the free energy $\mF[\phi,\epsilon_F]=\mE[\phi,\epsilon_F]-T\mS[\phi,\epsilon_F]$. This can be addressed by the Lagrange multiplier method. Define a new functional
\be\label{Yuk_grand}
\Omega[\phi,\epsilon_F,\mu]=\mF[\phi,\epsilon_F]-(\mQ[\phi,\epsilon_F]-Q)\mu,
\ee
where $\mu$ is the multiplier, and write down
\be\label{3_eqs}
\frac{\delta\Omega}{\delta\epsilon_F}=0,\quad \frac{\delta\Omega}{\delta\phi}=0,\quad \frac{\partial\Omega}{\partial\mu}=0.
\ee
The solution of the above three equations satisfies both the constraint and the minimization condition simultaneously.

The first equation of \Eq{3_eqs} yields
\be
\frac{\delta\Omega}{\delta\epsilon_F}=g_{\rm dof}\int\frac{\d^3p}{(2\pi)^3}(\epsilon_F-\mu)\left(\frac{\partial f}{\partial \epsilon_F}\right)=0.
\ee
Therefore $\epsilon_F(\x)\equiv\mu$ is a space-independent constant, defined as the chemical potential of the system.  Substituting this into the second equation, one obtains
\be\label{Yuk_eom_1}
\nabla^2\phi-\frac{\partial V}{\partial\phi}=g_{\rm dof}\,y\int\frac{\d^3p}{(2\pi)^3}\left(\frac{m}{\epsilon}\right)\frac{1}{e^{(\epsilon-\mu)/T}+1}.
\ee
The suppression factor $m/\epsilon$ implies the Yukawa interaction vanishes for massless fermions. This significantly differs from the case of a vector-mediated force, and the underlying reason is the difference between $\ave{\bar\chi\chi}$ and $\ave{\chi^\dagger\chi}$~\cite{Smirnov:2022sfo}.

For simplicity, hereafter we only consider the zero-temperature limit, in which $f(\x,\p)\to\Theta(\mu-\epsilon)$ with $\Theta$ the Heaviside step function. Then \Eq{Yuk_eom_1} is simplified to
\be\label{eom_yuk}
\nabla^2\phi=\frac{\partial V_{\rm eff}}{\partial\phi},
\ee
where the effective potential is
\be\label{Veff}
V_{\rm eff}(\phi)=V(\phi)+\frac{g_{\rm dof}}{16 \pi ^2}\left[\frac{\mu}{3}\sqrt{\mu^2-m^2} \left(5m^2-2\mu ^2\right)+m^4 \log\left(\frac{|m|}{\mu +\sqrt{\mu^2-m^2}}\right)\right],
\ee
where $m=y\phi$ as stated before. The first and second terms represent contributions from the bare scalar potential $V(\phi)$ and the fermion source $\ave{\bar\chi\chi}$, respectively. The shape of the fermion-induced potential is a symmetric valley located at $\phi\in(-\mu/y,\mu/y)$, indicating the Yukawa interaction is attractive, promoting the aggregation of fermions. This term is absent when $|\phi|\geqslant\mu/y$, or say, when $|m|\geqslant\mu$. This can be understood through the charge expression
\be\label{n_chi}
Q=\mQ[\phi,\epsilon_F]\Big|_{\epsilon_F=\mu}=\frac{g_{\rm dof}}{6\pi^2}\int\d^3x\left(\mu^2-m^2\right)^{3/2}\equiv\int\d^3x\, n,
\ee
where $n(\x)$ is the $\chi$ number density. Therefore, in the interior of the Fermi-ball, $|m|\leqslant\mu$ and $n>0$; at the boundary, $|m|=\mu$ and $n$ reaches 0. Outside the Fermi-ball, $|m|>\mu$ and $n\equiv0$, the second term in \Eq{Veff} vanishes.

\Eq{eom_yuk} is the basic EoM in this study. Apparently, $\phi(\x)\equiv w$ and hence $m(\x)\equiv yw=M$ is a trivial solution, corresponding to free fermion plane-waves within the true vacuum of $V(\phi)$. However, the EoM also allows for the existence of localized nontrivial $\phi(\x)$ distribution, which has a lower energy than the plane-wave solution. This can be demonstrated by rewriting the energy expression as 
\be\label{Esim}
E=\mE[\phi,\epsilon_F]\Big|_{\epsilon_F=\mu}=Q\mu+\int\d^3x\left[\frac12(\nabla\phi)^2+V_{\rm eff}(\phi)\right],
\ee
from which one can obtain
\be\label{dEdQ}
\frac{\delta E}{\delta Q}=\mu,
\ee
using the EoM and the $\mu$-dependency of $V_{\rm eff}$. Therefore, $\mu$ is the energy loss of a Fermi-ball after being extracted a $\chi$, and the process $(Q)\to \chi+(Q-1)$, i.e. a Fermi-ball with charge $Q$ evaporates to a free $\chi$ plus a smaller Fermi-ball with charge $(Q-1)$, releases an amount of energy $\Delta E=\mu-M$. If
\be\label{decay}
\mu<M,
\ee
then $\Delta E<0$, which means this process cannot happen spontaneously, and hence the Fermi-ball is stable against evaporation, called the quantum stability. Consequently, the nontrivial $\phi(\x)$ solution is energetically favored over the plane-wave solution. A second stability condition is
\be\label{split}
\frac{\delta^2E}{\delta Q^2}=\frac{\delta\mu}{\delta Q}<0,
\ee
which means the ball is stable agains splitting into smaller balls, also called the classical stability. Eqs. (\ref{decay}) and (\ref{split}) together imply a lower limit $Q_{\rm min}$ of the soliton charge, as will be demonstrated later.

\section{Resolving the equations of motion}\label{sec:solution}

\subsection{The algorithm and formation criterion}\label{subsec:algorithm}

The Fermi-ball is expected to be spherically symmetric, and hence all space-dependent functions such as $\phi$ and $m$ depend solely on $r$, the radial distance to the center of the ball. The EoM is then reduced to
\be\label{spherical_EoM}
\frac{\d^2\phi}{\d r^2}+\frac{2}{r}\frac{\d\phi}{\d r}=\frac{\partial V_{\rm eff}}{\partial\phi}.
\ee
The boundary conditions are stated as
\be\label{boundary_phi}
\frac{\d\phi}{\d r}\Big|_{r=0}=0,\quad \lim_{r\to\infty}\phi=w.
\ee
The first condition ensures the field is well-behaved at the center, while the second condition guarantees that $\phi$ is in the true vacuum of $V(\phi)$ outside the Fermi-ball. The radius of the ball $R$ is given by the condition $|m(R)|=\mu$ at which the $\chi$ number density is zero.

For the convenience of calculation, it is better to introduce the dimensionless variables
\be\label{yuk_dimensionless}
\xi=\mu\,r,\quad  u=\frac{m}{\mu}=\frac{y\phi}{\mu},
\ee
such that the EoM and the boundary conditions are rewritten as
\be\label{yuk_sim_eom2}
u''(\xi)+\frac2\xi u'(\xi)=\frac{\partial U_{\rm eff}}{\partial u};\quad u'(0)=0,\quad u(\infty)=u_w\equiv\frac{yw}{\mu}=\frac{M}{\mu},
\ee
where $U_{\rm eff}=(y^2/\mu^4)V_{\rm eff}$, while $u''$ and $u'$ represent $\d^2u/\d\xi^2$ and $\d u/\d\xi$, respectively. Note that $u_w>1$ is implied from the quantum stability condition $\mu<M$. The dimensionless effective potential is the difference between two functions,
\be\label{master}
U_{\rm eff}(u)=U(u)-F(u).
\ee
The former is a $\mu$-dependent potential $U=(y^2/\mu^4)V$ rescaling from the bare scalar potential. The latter is $\mu$-independent, representing $\ave{\bar\chi\chi}$ inside the soliton,
\be
F(u)=-g_{\rm dof}\frac{y^2}{16\pi^2}\left[\frac{5u^2-2}{3}\sqrt{1-u^2}+u^4\log\left(\frac{|u|}{1+\sqrt{1-u^2}}\right)\right],
\ee
for $u\in(-1,1)$ and it vanishes for $|u|\geqslant1$. The shape of $F(u)$ is a symmetric lump with $F(\pm1)=F'(\pm1)=F'(0)=0$ and $F(0)=g_{\rm dof}y^2/(24\pi^2)$.

\begin{figure}
\begin{center}
\includegraphics[scale=0.4]{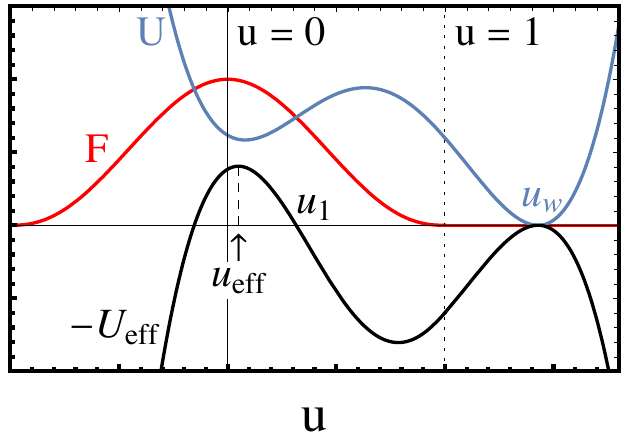}\qquad\qquad
\includegraphics[scale=0.4]{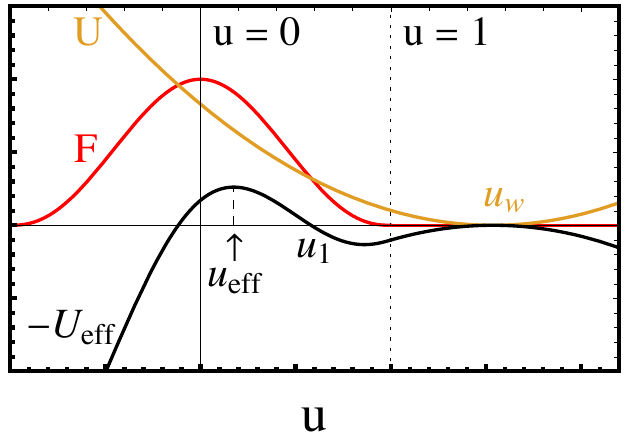}
\caption{Illustration of the functions $F(u)$, $U(u)$, and $-U_{\rm eff}(u)$ from different bare potential $V(\phi)$ inputs. $U(u)$ inherits the shape of $V(\phi)$, and can possess two ({\bf left}) or one ({\bf right}) vacuum. In either case, by appropriately choosing $\mu$, $-U_{\rm eff}(u)$ can have two hilltops, where the right one is at $u_w$, and the left one is at $u_{\rm eff}$. The root of $-U_{\rm eff}=0$ between $u_{\rm eff}$ and $u_w$ is denoted as $u_1$.}
\label{fig:Ueff}
\end{center}
\end{figure}

Treating $\xi$ as the effective ``time'' and $u$ as the effective ``position'', \Eq{yuk_sim_eom2} can be interpreted as the motion of a particle under a potential $-U_{\rm eff}$ and a time-dependent viscous damping force $-2u'/\xi$. The first boundary condition specifies the particle is dropped at rest, but $u(0)=u_0$ is undecided. We expect $u_0\in(0,1)$ because $\xi=0$ corresponds to the center of the soliton. For $u>1$, $-U_{\rm eff}$ has a local maximum at $u_w$ with $-U_{\rm eff}(u_w)=0$, denoted as the right-hilltop. The second boundary condition implies the particle can asymptotically climb up this hilltop at $\xi\to\infty$. Since $-U_{\rm eff}'(0)=F'(0)-U'(0)\geqslant0$, if $\mu$ is appropriately chosen, there could be another local maximum located at $u_{\rm eff}\in[0,1)$, with $-U'_{\rm eff}(u_{\rm eff})=0$, denoted as the left-hilltop. The shape of $-U_{\rm eff}(u)$ is illustrated in Fig.~\ref{fig:Ueff}.

When $-U_{\rm eff}$ has two hilltops and the left one is higher, i.e. $-U_{\rm eff}(u_{\rm eff})>-U_{\rm eff}(u_w)=0$, a solution to the EoM must exist. This can be understood through S. Coleman's overshoot-undershoot argument~\cite{Coleman:1977py}. Let $u_1\in(u_{\rm eff},u_w)$ be the zero point of $-U_{\rm eff}$. If $u_0\to u_{\rm eff}^+$, then the particle remains near the left-hilltop for a sufficiently long time before rolling down, such that the damping force is negligible. Then it will overshoot the right-hilltop due to mechanical energy conservation. Conversely, if $u_0=u_1$, then the particle does not have enough energy to ascend the right-hilltop due to the damping force. Therefore continuity ensures there must be a $u_0\in(u_{\rm eff},u_1)$ satisfying the boundary conditions.

The soliton formation condition of $-U_{\rm eff}(u)$ can be translated to that of $V_{\rm eff}(\phi)$: it should exhibit two vacua separated by a barrier. The \underline{false} vacuum originates from the bare potential $V(\phi)$ and is located at $w$, while the \underline{true} vacuum is situated at $w_{\rm eff}\equiv \mu u_{\rm eff}/y\in[0,w)$, including contributions from $\ave{\bar\chi\chi}$ inside the soliton. Fermi-balls reside near the true vacuum of $V_{\rm eff}(\phi)$, as previously discussed in the Introduction and depicted in the bottom panel of Fig.~\ref{fig:sketch}. In this sense, we observe no distinction between conventional ``solitons'' and ``bound states'', as both types of objects rely on the multi-vacuum structure of the effective potential $V_{\rm eff}(\phi)$. Bound states can be viewed as a subset of solitons, at least in the case of a large constituent number that mean field theory applies.

The above description not only proves the existence of the Fermi-ball, but also outlines the algorithm for evaluating its profile. Given the coupling $y$ and potential $V(\phi)$ in the model Lagrangian, follow these steps:
\begin{enumerate}
\item Choose a $\mu$, rescale $V(\phi)$ and $\phi$ to derive $U(u)$, $F(u)$ and then $U_{\rm eff}(u)$.
\item Check if the shape of $U_{\rm eff}$ is suited for soliton formation. If so, solve $u(\xi)$ from \Eq{yuk_sim_eom2} using the shooting method.
\item Transform $u(\xi)$ back to $\phi(r)$ to obtain the soliton profiles, including energy (mass) $E$, radius $R$, and charge $Q$.
\item Vary $\mu$ and evaluate a set of profiles, constructing the $\mu$-$Q$ mapping, which can be used to eliminate $\mu$ and express the soliton profiles as functions of the $Q$.
\end{enumerate}
By this procedure, we obtain the Fermi-ball profile for a specific particle physics model.

Step 4 warrants further elaboration. Since $\mu<M$ due to the quantum stability condition \Eq{decay}, and $\delta Q/\delta\mu<0$ due to the classical stability condition \Eq{split}, $Q$ must have a lower limit $Q_{\rm min}$. This can be understood via the shape of the function $Q(\mu)$. If $Q(\mu)$ is monotonic decreasing for all $\mu< M$, then $Q$ reaches its minimum at $\mu=M$, at which $Q_{\rm min}=Q_S$ is the quantum stability charge. If, however, $Q(\mu)$ has a minimum at $\mu=\mu_C<M$, then $Q_{\rm min}=Q_C\equiv Q(\mu_C)$ is given by the classical stability~\cite{Koeppel:1985tt}, and $\mu_C$ is the maximal value that $\mu$ can take for soliton formation. Fermi-balls with $Q<Q_{\rm min}$ either evaporate to free $\chi$ particles or split to smaller balls, and eventually disappear. When the stability conditions are satisfied, as $Q$ increases, $\mu$ decreases, resulting in a more stable soliton. However, reducing $\mu$ enlarges $U=(y^2/\mu^4)V$, and hence $-U_{\rm eff}$ decreases. This sets a lower limit $\mu_{\rm min}$, at which $U(u)$ and $F(u)$ are tangent at $u_{\rm eff}$, leading to the left-hilltop height $-U_{\rm eff}(u_{\rm eff})=0$. When $\mu<\mu_{\rm min}$, $U(u)$ surpasses $F(u)$ across the entire range $u\in[-1,1]$ and hence the left-hilltop is lower than the right-hilltop or it even does not exist. As a consequence, the Fermi-balls cannot form. When $\mu\to\mu_{\rm min}^+$, both $Q$ and $R$ diverge, yet the ratio $Q/R^3$ approaches a finite constant, dubbed the saturation limit.

\subsection{Two important analytical limits}

Calculating the Fermi-ball profile involves solving highly nonlinear differential equations, as detailed in the previous subsection. However, under certain limits, the profile can be analytically derived. Here, I will discuss two significant cases: the saturation limit and the Yukawa attraction limit, demonstrating that the results can be reduced to well-known analytical formulae found in the literature.

\subsubsection*{The saturation limit}

This is the case implicitly assumed in most phenomenological studies. As $\mu\to\mu_{\rm min}^+$, $U(u)$ tends to be tangent with $F(u)$ at $u\approx u_{\rm eff}$, leading to $-U_{\rm eff}(u_{\rm eff})\to 0^+$ and $u_0\to u_{\rm eff}^+$. In this case, the particle stays close to the left-hilltop for a long time until $\xi\approx\xi_R$ when the damping force is negligible, and it rapidly rolls down to the right-hilltop. Therefore, for $\xi\ll\xi_R$, $u\approx u_{\rm eff}$; for $\xi\gg\xi_R$, $u\approx u_w$; while for $\xi\sim\xi_R$, $u'\approx\sqrt{2U_{\rm eff}}$, as inferred from the EoM without the damping term. This implies the scalar field has a nearly constant value $\phi(r)\approx w_{\rm eff}$ inside the soliton and $\approx w$ outside the soliton, with the boundary located at $R=\xi_R/\mu$, characterized by a thin wall. Therefore, this is also called the ``thin-wall'' limit of the EoM, which occurs at $Q\to\infty$ and the profile can be analytically resolved~\cite{Coleman:1977py}.

At this limit, the $\chi$ particles in the soliton distribute uniformly. The charge integration \Eq{n_chi} can be analytically evaluated,
\be
Q\xrightarrow[]{\mu\to\mu_{\rm min}^+}\frac{4\pi}{3}R^3\frac{g_{\rm dof}}{6\pi^2}\left(\mu^2-M_{\rm eff}^2\right)^{3/2},
\ee
where $M_{\rm eff}=yw_{\rm eff}$ is the $\chi$ mass inside the soliton. Then we can get
\be\label{saturation_mu}
\mu_{\rm min}=\lim_{Q\to\infty}\sqrt{\left(\frac{9\pi Q}{2g_{\rm dof} R^3}\right)^{2/3}+M_{\rm eff}^2},
\ee
and hence the fermion number density $\propto Q/R^3$ is finite. The existence condition of soliton requires $\mu_{\rm min}<M$, to wit
\be
\lim_{Q\to\infty}\left(\frac{9\pi Q}{2g_{\rm dof} R^3}\right)^{2/3}<M^2-M_{\rm eff}^2.
\ee
This means the trapped fermions do not have sufficient kinetic energy to overcome the mass gap between the two vacua, a frequently adopted statement in the literature.

In the saturation limit, the energy integration \Eq{Esim} can be categorized into three ranges: interior ($\xi<\xi_R$), exterior ($\xi>\xi_R$), and surface ($\xi\sim\xi_R$) of the Fermi-ball. Using the expression of $\mu_{\rm min}$, we find $E\to E_{\rm kin}+E_{\rm vac}+E_{\rm surf}$, where the Fermi-gas kinetic term
\be
E_{\rm kin}=\frac{3\pi}{4}\left(\frac{3}{2\pi}\right)^{2/3}\left(\frac{2}{g_{\rm dof}}\right)^{1/3}\frac{Q^{4/3}}{R}K(\zeta),
\ee
with $\zeta=2(g_{\rm dof}/2)^{1/3}M_{\rm eff} R/(18\pi Q)^{1/3}= u_{\rm eff}/\sqrt{1-u_{\rm eff}^2}$, and the special function is
\be\label{K}
K(\zeta)=\left(1+\frac{\zeta^2}{2}\right)\sqrt{1+\zeta^2}+\frac{\zeta^4}{2}\log\left(\frac{\zeta}{1+\sqrt{1+\zeta^2}}\right)\approx\begin{dcases}~1,&\zeta\ll1;\\
~\frac{4\zeta}{3}+\frac{2}{5\zeta},&\zeta\gg1.
\end{dcases}
\ee
The vacuum energy term $E_{\rm vac}=(4\pi R^3/3)V_0$, where $V_0=V(w_{\rm eff})$ is the bare scalar potential energy. The surface tension term is $E_{\rm surf}=4\pi R^2\sigma$ with
\be
\sigma=\int_{w_{\rm eff}}^{w}\d\phi\sqrt{2V_{\rm eff}(\phi)}.
\ee
Typically, the surface term is significantly smaller than the volume term, leading to the neglect of $E_{\rm surf}$, particularly when Fermi-balls exhibit macroscopic sizes. However, in scenarios where $V(\phi)$ has near-degenerate vacua, the surface term becomes crucial.

\begin{table}\small\renewcommand\arraystretch{1.5}\centering
\begin{tabular}{|c|c|c|}\hline
$\mu\to\mu_{\rm min}^+$ & Relativistic constituents $\zeta\ll1$ & Non-Relativistic constituents $\zeta\gg1$ \\ \hline
$E_{\rm kin}$ & $\frac{3\pi}{4}\left(\frac{3}{2\pi}\right)^{2/3}\left(\frac{2}{g_{\rm dof}}\right)^{1/3}\frac{Q^{4/3}}{R}$ & $QM_{\rm eff}+\frac{2\pi^{2/3}}{15M_{\rm eff}}\left(\frac{9}{4}\right)^{5/3}\left(\frac{2}{g_{\rm dof}}\right)^{2/3}\frac{Q^{5/3}}{R^2}$ \\ \hline
$E$ & $Q\left[\left(\frac{2}{g_{\rm dof}}\right)12\pi^2V_0\right]^{1/4}$ & $QM_{\rm eff}\left[1+\frac{\left(15\pi^2\right)^{2/5}}{2}\left(\frac{2}{g_{\rm dof}}\right)^{2/5}\left(\frac{V_0^{1/4}}{M_{\rm eff}}\right)^{8/5}\right]$ \\ \hline
$R$ & $Q^{1/3}\left[\frac{3}{16}\left(\frac{3}{2\pi}\right)^{2/3}\left(\frac{2}{g_{\rm dof}}\right)^{1/3}\frac{1}{V_0}\right]^{1/4}$ & $Q^{1/3}\frac{\left(3^7/\pi\right)^{1/15}}{2^{2/3}5^{1/5}}\left(\frac{2}{g_{\rm dof}}\right)^{2/15}\left(\frac{1}{M_{\rm eff}V_0}\right)^{1/5}$ \\ \hline
\end{tabular}
\caption{Fermi-ball profiles at the saturation limit, derived neglecting the surface tension term.}
\label{tab:saturation}
\end{table}

Given charge $Q$, the radius $R$ can be determined by minimizing $E$, i.e. $\d E/\d R=0$.\footnote{Strictly speaking, the radius is given by minimizing the free energy; however $T=0$ is considered here, thus free energy equals to energy.} This is the approach commonly used in most studies to derive soliton profiles. The results for relativistic and non-relativistic cases are listed in Table~\ref{tab:saturation} and are consistent with those found in the literature. Two points are worth mentioning. First, Fermi-balls are in the true vacuum $w_{\rm eff}$ of the effective potential $V_{\rm eff}$ rather than the false vacuum $w'$ (if exists) of the bare potential $V$, although the difference is expected to be small. Second, while the scaling $E\propto Q$ is obtained, it does not mean $\delta^2E/\delta Q^2=0$ and the stability against splitting is violated. In fact, the formulae in Table~\ref{tab:saturation} are obtained neglecting the surface tension term; after including $E_{\rm surf}\propto Q^{2/3}$, $\delta^2E/\delta Q^2<0$ still holds.

It is important to note that the condition of the saturation limit is not universally met in every concrete phenomenological model. Specifically, if Fermi-balls form very shortly after the Big Bang (or equivalently, at very high scales), when the total number of $\chi$ particles contained within a horizon volume is small, the charge $Q$ accumulated in a ball must also be small, potentially falling outside the scope of the saturation limit. In such cases, it is necessary to numerically solve the EoMs to obtain accurate profiles.

\subsubsection*{The Yukawa attraction limit}

This is the case when the Fermi-ball can be viewed as non-relativistic fermions being bound by the classical Yukawa attractive force. Assume $u_{\rm eff}\ll u_0,~u_1,~u_w\sim1$, and in the neighborhood of $u_w$ the potential can be approximated as
\be\label{Vphiu01}
V(\phi)\xrightarrow[]{u_0\sim1}\frac{m'^2}{2}(\phi-w)^2=\frac{m'^2}{2}\left(\phi-\frac{M}{y}\right)^2;\quad
U(u)\xrightarrow[]{u_0\sim1}\frac{m'^2}{2\mu^2}(u-1)^2,
\ee
where $m'$ is the scalar mass outside the Fermi-ball. As $u_w=M/\mu$ by definition, $u_w\sim1$ means $\mu$ is not much smaller than $M$, or say $Q$ is not far away from $Q_{\rm min}=Q_S$. Define $\varphi=\phi-w$, the Lagrangian \Eq{Yuk_bound_L} is rewritten as
\be
\mL\xrightarrow[]{u_0\sim1}\bar\chi(i\slashed{\partial}-M)\chi+\frac{1}{2}\partial_\mu\varphi\partial^\mu\varphi-\frac{1}{2}m'^2\varphi^2,
\ee
which describes the motion of fermion $\chi$ with mass $M$ interacting via the Yukawa force mediated by the scalar boson $\varphi$. The energy is
\be
E\xrightarrow[]{u_0\sim1}\int\d^3x\left[\bar\chi\left(-i\vec{\gamma}\cdot\nabla+M\right)\chi+\left(y\varphi\bar\chi\chi+\frac12(\nabla\varphi)^2+\frac{1}{2}m'^2\varphi^2\right)\right],
\ee
in the form of $E_{\rm kin}+E_{\rm Yuk}$, the summation of Fermi-gas kinetic energy and Yukawa binding energy. Since $u_0\sim1$, the scalar field as wall as fermion number density in the Fermi-ball can be approximately treated as uniform, and hence $E_{\rm kin}$ is given by replacing the $M_{\rm eff}$ with $M$ in the $E_{\rm kin}$ expression in the $\zeta\gg1$ case of Table~\ref{tab:saturation}.

The Yukawa energy is obtained via integration by part and applying the EoMs,
\be\label{Eyukawa}
E_{\rm Yuk}=\int\d^3x\frac{1}{2}y\varphi\ave{\bar\chi\chi}.
\ee
In the non-relativistic limit, $\ave{\bar\chi\chi}\approx\ave{\chi^\dagger\chi}=n(\x)$, which is the $\chi$ number density; therefore, the $(m/\epsilon)$ factor in \Eq{Yuk_eom_1} can be dropped, and the EoM reduces to $\nabla^2\varphi-m'^2\varphi^2\approx yn$, which can be solved by the Green's function method
\be
\varphi(\x)\approx-\frac{y}{4\pi}\int\d^3x'n(\x')\frac{e^{-m'|\x-\x'|}}{|\x-\x'|}.
\ee
Substituting this to \Eq{Eyukawa} and adopting $n\approx 3Q/(4\pi R^3)$, one obtains
\be\label{EYUNR}
E_{\rm Yuk}\approx-\frac{y^2}{8\pi}\frac{9Q^2}{16\pi^2R^6}\int\d^3x\int\d^3x'\frac{e^{-m'|\x-\x'|}}{|\x-\x'|}=-\frac{3y^2Q^2}{20\pi R}f_y\left(\frac{1}{m'R}\right),
\ee
with the special function
\be
f_y(\eta)=\frac52\eta^2\left[1+\frac32\eta(\eta^2-1)-\frac32\eta(\eta+1)^2e^{-2/\eta}\right],
\ee
satisfying $f_y(0)=0$ and $f_y(\infty)=1$. $E_{\rm Yuk}<0$ because Yukawa force is attractive.

The Fermi-ball profile can be obtained by minimizing $E\to E_{\rm kin}+E_{\rm Yuk}$ with respect to the radius $R$. $m'\to0$ is called the Coulomb limit, as the range of force $m'^{-1}\to\infty$. In this regime, the Fermi-ball profile can be analytically obtained
\be\label{NRlimit}
E\xrightarrow[m'\to0]{u_0\sim1}QM\left[1-\left(\frac{g_{\rm dof}^2}{6\pi^2}\right)^{1/3}\frac{y^4Q^{4/3}}{80\pi^2}\right],\quad
R\xrightarrow[m'\to0]{u_0\sim1}\left(\frac{2}{g_{\rm dof}}\right)^{2/3}\frac{4\pi(9\pi/4)^{2/3}}{y^2MQ^{1/3}},
\ee
showing a scaling behavior of $R\sim Q^{-1/3}$, thus the Fermi-ball is more compact at larger charges, because the attraction binds the fermions more tightly. But this scaling does not hold when $Q\to\infty$, as this corresponds to the saturation limit where $R\propto Q^{1/3}$. Hence, one expects $R$ first decreases and then increases with $Q$. This feature is known in studies of fermion bound states~\cite{Stephenson:1996qj,Wise:2014jva,Wise:2014ola,Gresham:2017zqi,Gresham:2018anj,Smirnov:2022sfo}, where $V(\phi)\sim m'^2(\phi-w)^2/2$ is assumed and hence \Eq{Vphiu01} holds across the entire range of $\phi$ or $u$. The behavior of $R$ is traditionally interpreted as follows: as $Q\to\infty$, constituent $\chi$'s are compressed into the relativistic regime, leading to the suppression of the Yukawa interaction by the factor $m/\epsilon$ factor in \Eq{Yuk_eom_1}, resulting in the breakdown of the Yukawa attraction condition and the $R\sim Q^{-1/3}$ scaling. However, this research offers another interpretation.

As discussed around \Eq{K} and listed in Table~\ref{tab:saturation}, whether the $\chi$ particles are relativistic depends on the parameter $\zeta=u_{\rm eff}/\sqrt{1-u_{\rm eff}^2}$. In conventional bound state studies, $V(\phi)$ is a quadratic potential, resulting in a small $u_{\rm eff}$ when $\mu\to\mu_{\rm min}^+$, leading to relativistic $\chi$'s when $Q\to\infty$. However, if we switch to a shallower potential, e.g. $V(\phi)\sim\lambda(\phi-w)^4$, then $u_{\rm eff}$ could be $\sim1$ even when $\mu\to\mu_{\rm min}^+$, implying the $\chi$'s inside the soliton are still non-relativistic at $Q\to\infty$, and the Yukawa interaction is not suppressed. But even in this case, we obtain $R\to Q^{1/3}$ for $Q\to\infty$. The reason is that the crucial condition for the Yukawa attraction limit is not the smallness of $u_{\rm eff}$ but the separation between $u_{\rm eff}$ and $u_1$. When $Q$ increases, $\mu$ decreases accordingly, and hence $U(u)$ is enhanced and the left-hilltop of $-U_{\rm eff}$ is lowered, making $u_1$ shift leftward while $u_{\rm eff}$ shift rightward. Eventually, when $\mu\to\mu_{\rm min}^+$, $u_{\rm eff}\to u_1$, and the releasing point $u_0$ is unavoidably close to the left-hilltop $u_{\rm eff}$, which makes it stays for a long time before rolling, and hence $R\to\infty$ and the $R\propto Q^{1/3}$ scaling is obtained.

\section{Two illustrative examples}\label{sec:numerical}

\begin{figure}
\begin{center}
\includegraphics[scale=0.35]{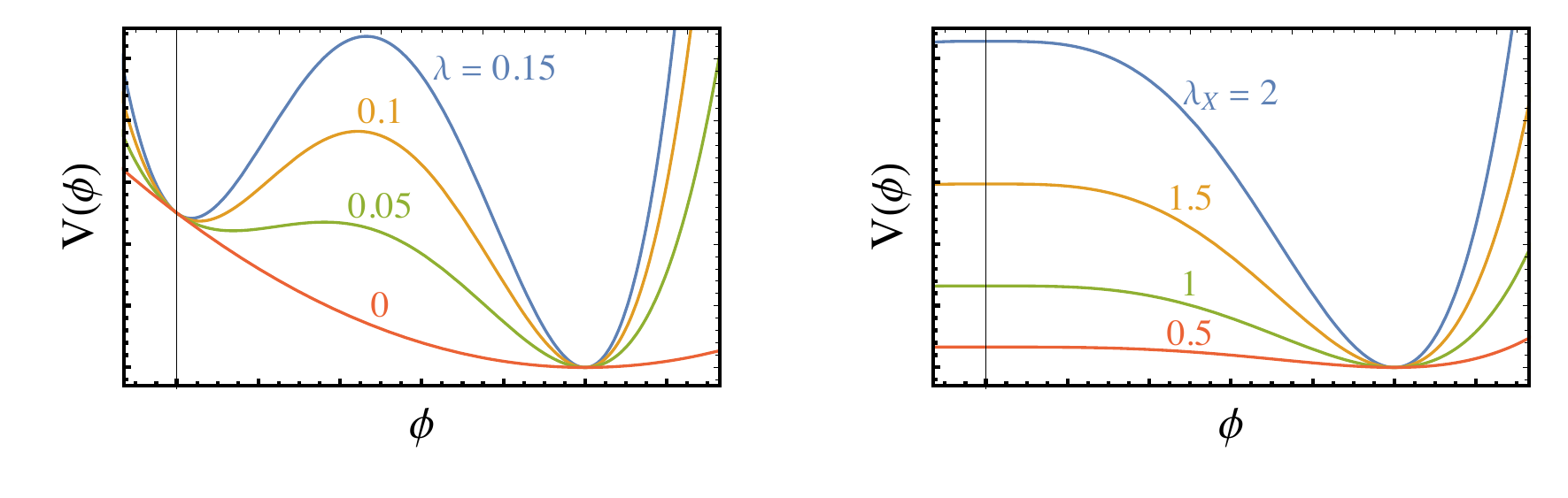}
\caption{\textbf{Left}: polynomial potential \Eq{toy} with $w=1$ TeV, $a=50$ GeV, and different $\lambda$'s. \textbf{Right}: logarithmic potential \Eq{cc} with $w=1$ TeV and different $\lambda_X$'s. The vertical black lines are at $\phi=0$. See the text for the definition of the potentials.}
\label{fig:V}
\end{center}
\end{figure}

To demonstrate the algorithm and the features of the Fermi-ball, this section provides two numerical solution examples, namely the polynomial and logarithmic potentials.

\subsection{The polynomial potential: a combined discussion of conventional ``solitons'' and ``bound states''}

The scalar potential is
\be\label{toy}
V(\phi)=\frac{a^2}{2}(\phi-w)^2\left(1+\frac{\lambda\phi^2}{2a^2}\right),
\ee
where $a>0$ and $\lambda\geqslant0$. The potential has a global minimum at $w$, at which the scalar boson mass is $m'=\sqrt{a^2+\lambda w^2/2}$. If $\lambda>16a^2/w^2$, a local minimum
\be
w'=\frac{w}{4}\left(1-\sqrt{1-\frac{16a^2}{\lambda w^2}}\right),
\ee
exists in $\phi\in(0,w)$. This model is chosen as an illustrative example due to its ability to transition smoothly from a double-vacuum structure to a single-vacuum structure by adjusting the parameter $\lambda$, making it well-suited for the discussions at hand. We consider four distinct values of $\lambda=0.15$, 0.1, 0.05 and $0$, where the first three represent the ``traditional soliton'' scenario, and the last one represents the ``traditional bound state'' scenario of $V(\phi)$. Other parameters are set as $y=1$, $a/w = 0.05$, and the results are irrelevant to $w$ if all observables are rescaled to be dimensionless using $w$ and $a$. The shapes of $V(\phi)$ are depicted in the left panel of Fig.~\ref{fig:V}.

\begin{figure}
\begin{center}
\includegraphics[width=\textwidth]{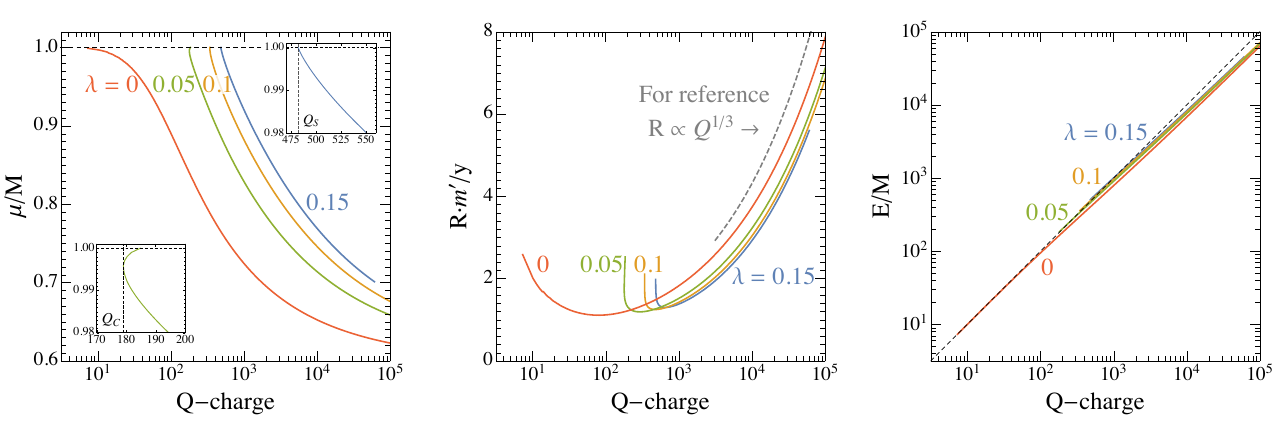}
\caption{The Fermi-ball profiles as functions of $Q$, for different $\lambda$'s in the polynomial potential of \Eq{toy}. {\bf Left}: the chemical potential $\mu$ over the free fermion mass $M$. {\bf Middle}: the radius over $y/m'$, where $m'$ is the scalar mass outside the ball. {\bf Right}: the energy (mass) over $M$.}
\label{fig:scan}
\end{center}
\end{figure}

The algorithm outlined in section~\ref{subsec:algorithm} is implemented for each $\lambda$ benchmark. $\mu_{\rm min}$ is determined by the tangency condition between $U(u)$ and $F(u)$. Subsequently, we scan over $\mu\in(\mu_{\rm min},M)$ to get the profiles such as charge $Q$, energy (mass) $E$, radius $R$, etc. Using the $\mu$-$Q$ correspondence, one can express all profiles as functions of $Q$, with some exemplified in Fig.~\ref{fig:scan}. The left panel is the curve for chemical potential $\mu$ relative to the free fermion mass $M$, and it must be smaller than 1 from the quantum stability condition. As discussed at the end of section~\ref{subsec:algorithm}, there exists a minimal charge $Q_{\rm min}$, below which no stable soliton forms. This can be seen clearly in the left panel of Fig.~\ref{fig:scan}: for $\lambda=0$ or 0.15, $Q_{\rm min}$ is given by the quantum stability charge $Q_S$; while for $\lambda=0.1$ or 0.05, $Q_{\rm min}$ is given by the classical stability charge $Q_C$. The two zoomed-in sub-figures show the cases of $\lambda=0.15$ and 0.05, where $Q_{\rm min}$'s are given by $Q_S$ and $Q_C$, respectively. We have checked that $\mu=\delta E/\delta Q$ holds very well for the numerical results. Also note $\mu$ decreases with $Q$ in the parameter space viable for soliton formation; Fermi-balls are more stable with greater accumulated charges.

The radius profile in the middle panel of Fig.~\ref{fig:scan} exhibits a trend where $R$ initially decreases and then increases with increasing $Q$, as previously analyzed. The initial decrease in $R$ at small $Q$ is caused by the Yukawa attraction limit, although it does not precisely follow an $R \propto Q^{-1/3}$ scaling due to the non-negligible $m'R$ in this scenario. As $Q$ rises, the Fermi-ball approaches the saturation limit, where $R \propto Q^{1/3}$. The transition point between these two regimes, denoted as $Q_{\rm cri}$, can be roughly estimated by matching the fermion number density in \Eq{NRlimit} to that in the $\zeta\ll1$ case of Table~\ref{tab:saturation}, yielding
\be\label{Qcri}
Q_{\rm cri}\sim 24\sqrt{3}\pi^3\left(\frac{2}{g_{\rm dof}}\right)\frac{\sqrt{V(0)}}{y^3M^2}.
\ee
For the chosen benchmarks, $Q_{\rm cri}\sim\mO(100)$ is obtained, consistent with the numerical findings shown in the figure. The right panel of Fig.~\ref{fig:scan} depicts the energy (mass) profiles. Given that both the non-relativistic and saturation limits yield $E\propto Q$, the shape of the curves remains relatively stable across the entire $Q$ range.

\begin{figure}
\begin{center}
\includegraphics[scale=0.32]{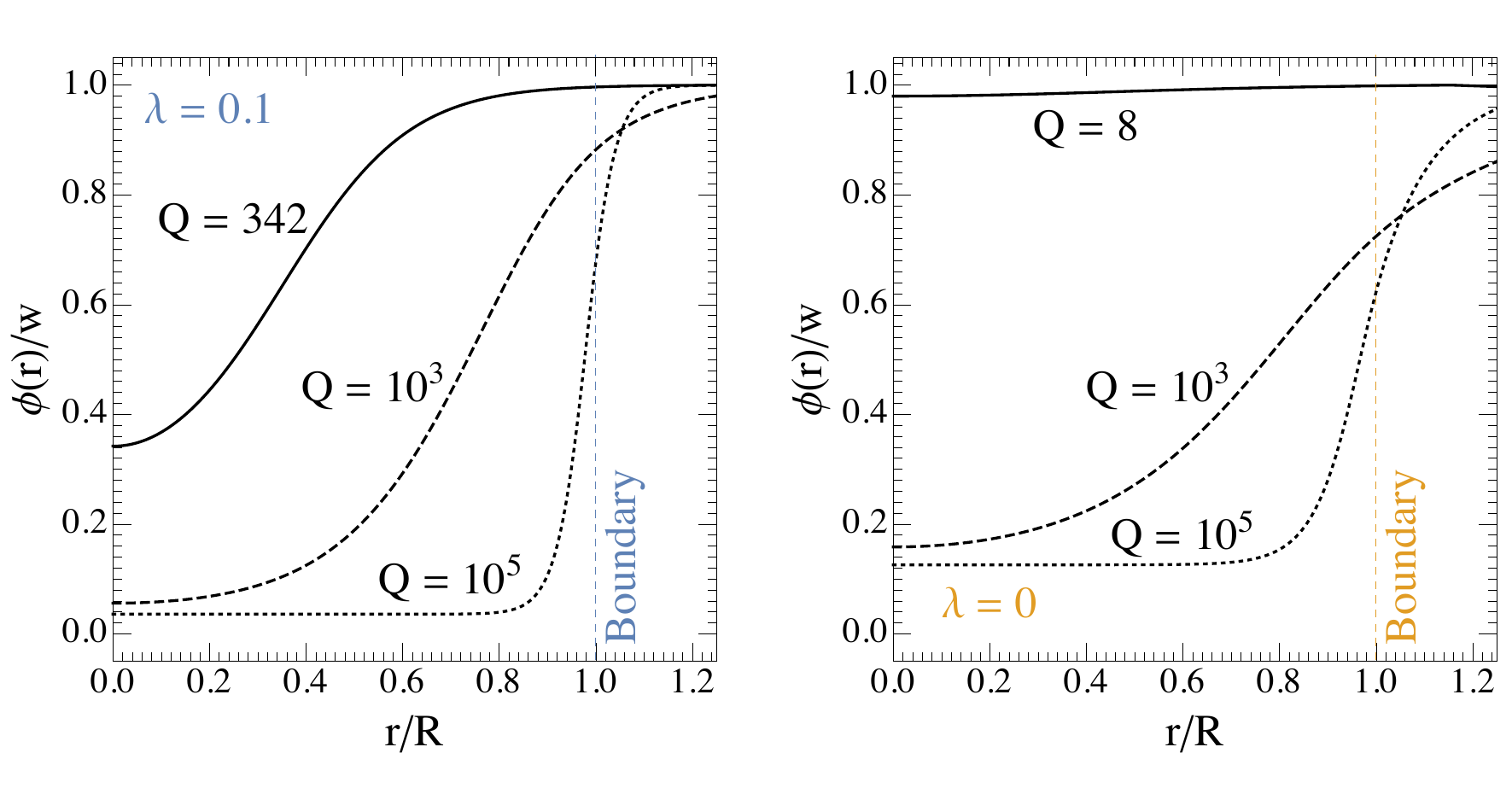}
\caption{The scalar distribution $\phi(r)$ of a single Fermi-ball at different $Q$'s for $\lambda=0.1$ ({\bf left}) and $\lambda=0$ ({\bf right}) for the polynomial potential of \Eq{toy}. The horizontal axis is rescaled by the soliton radius $R$.}
\label{fig:profiles}
\end{center}
\end{figure}

Fig.~\ref{fig:profiles} displays examples of the $\phi(r)$ distribution profiles of a single Fermi-ball for various $Q$ charges. The radius $R$ at different $Q$ values can exhibit significant differences, and hence for comparison we normalize the radial distance by $R$, where $r/R = 1$ corresponds to the soliton boundary. Notably, as $r/R\gg1$, $\phi(r)$ approaches $w$, indicating the $V(\phi)$ true vacuum in the exterior of the Fermi-ball. Conversely, the interior of the ball always has a lower $\phi$ field value (and hence a lower $\chi$ mass), with $\phi(0)$ tending toward $w_{\rm eff}$ as $Q\to\infty$ in the saturation limit. In that case, the $\phi$ distribution tends to become uniform within the ball, yielding a uniform fermion number density -- a common assumption in phenomenological studies, as known as the ``thin wall approximation''. However, for small $Q$, deviations from this approximation are substantial, emphasizing the necessity of accurately solving the EoMs.

Gradually decreasing $\lambda$ to 0 while keeping $Q$ as a constant results in the disappearance of the barrier and a transition in the shape of the bare potential $V(\phi)$ from double- to single-vacuum. During this procedure, we observe a smooth transition in the $\phi(r)$ profile, as can be seen by comparing the curves between the left and right panels in Fig.~\ref{fig:profiles} at the same $Q$. This suggests that there is no fundamental distinction between conventional fermionic solitons and fermion bound states; rather, the latter can be regarded as a subset of the former.

\subsection{The classically conformal potential: a non-polynomial case}

\begin{figure}
\begin{center}
\includegraphics[width=\textwidth]{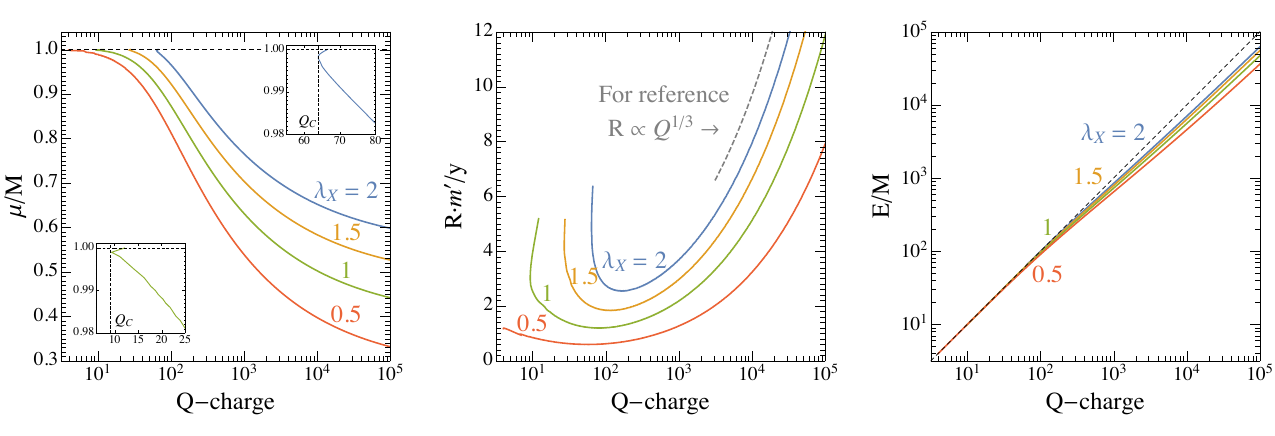}
\caption{The Fermi-ball profiles as functions of $Q$, for different $\lambda_X$'s in the logarithmic potential of \Eq{cc}. {\bf Left}: the chemical potential $\mu$ over the free fermion mass $M$. {\bf Middle}: the radius over $y/m'$, where $m'$ is the scalar mass outside the ball. {\bf Right}: the energy (mass) over $M$.}
\label{fig:scan_2}
\end{center}
\end{figure}

Motivated by the classically conformal theories~\cite{Iso:2009ss,Iso:2009nw,Chun:2013soa}, the potential is logarithmic and can be parametrized as
\be\label{cc}
V(\phi)=\frac{\lambda_X^2}{128\pi^2}\phi^4\left(\log\frac{\phi}{w}-\frac14\right),
\ee
where $\lambda_X>0$ characterizes the coupling of $\phi$ to some bosonic degrees of freedom in the model. This potential has only one single vacuum at $\phi=w$ whose depth can be tuned via varying $\lambda_X$, as shown in the right panel of Fig.~\ref{fig:V}. The scalar mass is $m'=\lambda_Xw/(4\sqrt{2}\pi)$ at vacuum.

The logarithmic potential $V$ never exhibits a barrier separating two local minima, yet we still find Fermi-ball solutions thanks to the inclusion of the $\ave{\bar\chi\chi}$ term in the EoM, which generates a second minima for $V_{\rm eff}$. We fix $w=1$ TeV adopt $\lambda_X=2$, 1.5, 1, and 0.5 as benchmark values, and vary the chemical potential $\mu$ to derive the Fermi-ball profiles. The results are shown in Fig.~\ref{fig:scan_2}, which show characteristics similar to the case of the polynomial potential. The left panel demonstrates the decreasing $\mu$ with $Q$ in the parameter space allowing for soliton formation, and the existence of minimal charge $Q_{\rm min}$. The first negative and then positive correlation between radius $R$ and the charge $Q$ is clearly shown in the middle panel, with the saturation limit $R\propto Q^{1/3}$ being achieved at large $Q$. Finally, the right panel plots the energy (mass) dependence of $Q$, and the $E/M$ curves are below $Q$ (dashed black line) because of the quantum stability. Those common features confirm our general analysis based on the EoM.

\section{Phenomenology: formation, evolution, and signals}\label{sec:pheno}

The preceding sections have demonstrated the existence of Fermi-ball solutions and outlined the methodology for obtaining them. However, the actual production of these solitons in the Universe is another topic, which entails the nontrivial dynamics of the particle model. This section will explore various formation mechanisms and examine the subsequent fate of Fermi-balls post-production, including whether they can collapse to primordial black holes (PBHs). The experimental signals are also discussed.

\subsection*{Formation}

A straightforward and intuitive mechanism is the direct fusion of free $\chi$ particles to a Fermi-ball: if the net charge $Q$ within a region with size $R$ exceeds the minimal charge $Q_{\rm min}$ for soliton formation, free $\chi$ particles can fuse to a Fermi-ball, similar to the fusion of free scalar particles to a Q-ball~\cite{Kusenko:1997hj}, although detailed studies on a general $V(\phi)$ are lacking in the fermion case (the case of the conventional bound states has been investigated~\cite{Gresham:2017cvl}). Apparently, the probability of this process is highly suppressed if $Q_{\rm min}\gg1$, thus it can be important only for the models with order-one $Q_{\rm min}$. Considering the hot and dense plasma of the early Universe, where the bare potential $V(\phi)$ is replaced by the finite temperature potential $V(\phi,T)$, and hence $Q_{\rm min}$ also becomes temperature-dependent, the feasibility of the fusion process might be relaxed to requiring $Q_{\rm min}(T)\sim\mO(1)$ during some period of the cosmic thermal history.

Another plausible and extensively studied scenario involves a first-order phase transition (FOPT). While the Standard Model (SM) of particle physics lacks a FOPT in the thermal history of the Universe, many models beyond the SM do exhibit this phenomenon. In such models, the finite temperature $V(\phi,T)$ has two distinct vacua separated by a barrier. Consistent with the notation in section~\ref{sec:framework}, we denote the true vacuum as $w$ and the false vacuum as $w'$. The Universe is initially in $w'$, and decays to $w$, triggering a cosmic FOPT at temperature $T_*$~\cite{Mazumdar:2018dfl,Athron:2023xlk}. This results in true vacuum bubble nucleation and expansion, during which the $\chi$ particles have masses $M'=yw'$ and $M=yw$ in the false and true vacua, respectively. If $(M-M')/T_*\gtrsim\mO(10)$, the fermions do not have sufficient kinetic energy to penetrate into the true vacuum. As a result, they remain trapped in the false vacuum, eventually forming Fermi-balls.

\begin{table}\footnotesize\renewcommand\arraystretch{1.5}\centering
\begin{tabular}{|c|c|c|}\hline
 & Preexisting $\chi$-asymmetry & Thermal fluctuations \\ \hline
$\ave{Q}$ & $10^{47}\times v_w^3\left(\frac{Y}{10^{-10}}\right)\left(\frac{\rm GeV}{T_*}\right)^3\left(\frac{H_*}{\beta}\right)^3$ & $10^{27}\times v_w^{3/2}\left(\frac{\rm GeV}{T_*}\right)^{3/2}\left(\frac{H_*}{\beta}\right)^{3/2}$ \\ \hline
$\ave{E}$ & $10^{24}~{\rm g}\times v_w^3\left(\frac{Y}{10^{-10}}\right)\left(\frac{\rm GeV}{T_*}\right)^2\left(\frac{H_*}{\beta}\right)^3\alpha^{1/4}$ & $10^4~{\rm g}\times v_w^{3/2}\left(\frac{\rm GeV}{T_*}\right)^{1/2}\left(\frac{H_*}{\beta}\right)^{3/2}\alpha^{1/4}$ \\ \hline
$\ave{R}$ & $10~{\rm cm}\times v_w\left(\frac{Y}{10^{-10}}\right)^{1/3}\left(\frac{\rm GeV}{T_*}\right)^2\left(\frac{H_*}{\beta}\right)\alpha^{-1/4}$ & $10^{-6}~{\rm cm}\times v_w^{1/2}\left(\frac{\rm GeV}{T_*}\right)^{3/2}\left(\frac{H_*}{\beta}\right)^{1/2}\alpha^{-1/4}$ \\ \hline
$f_{\rm dm}$ & $\left(\frac{Y}{10^{-10}}\right)\left(\frac{T_*}{\rm GeV}\right)\alpha^{1/4}$ & $10^{-20}\times v_w^{-3/2}\left(\frac{T_*}{\rm GeV}\right)^5\left(\frac{\beta}{H_*}\right)^{3/2}\alpha^{1/4}$ \\ \hline
\end{tabular}
\caption{Estimate of the Fermi-ball profiles induced by FOPTs at the relativistic saturation limit. The FOPT parameters are $T_*$ (transition temperature), $\alpha$ (latent heat over the radiation energy), $\beta/H_*$ (Hubble time over the transition duration), and $v_w$ (bubble expansion velocity)~\cite{Caprini:2019egz}. $f_{\rm dm}$ is the Fermi-ball fraction of dark matter today.}
\label{tab:fopt_fbs}
\end{table}

Fermi-balls must exclusively consist of either $\chi$ or $\bar\chi$ particles to prevent instability from $\chi\bar\chi$ annihilation. This requires a number density asymmetry between the trapped $\chi$'s and $\bar\chi$'s, such that one type survives the annihilation at the final stage of the false vacuum remnant shrinking. This asymmetry can arise from a preexisting $\chi$-asymmetry $Y=(n_\chi-n_{\bar\chi})/s$ with $s$ being the entropy density~\cite{Hong:2020est}, or through thermal fluctuations~\cite{Asadi:2021yml}. The saturation limit analytical formulae listed in Table~\ref{tab:fopt_fbs} can be employed to estimate the profiles, and details are provided in Appendix~\ref{app:FOPT}. Note that $Q$, $E$, and $R$ are all negatively correlated to $T_*$, implying that Fermi-balls formed at higher scales (earlier in time) are lighter and smaller, and may fail to meet the saturation condition $Q\gg Q_{\rm cri}$, necessitating refined calculations. Trapping fermions in the false vacuum in a FOPT to form solitons as macroscopic dark matter is an extensively studied scenario~\cite{Bai:2018vik,Bai:2018dxf,Hong:2020est,Marfatia:2021twj,Gross:2021qgx}.

Fermi-balls may also arise from second-order phase transitions through scalar field fragmentations, akin to the formation of Q-balls in the evolution of the Affleck-Dine field~\cite{Kusenko:1997si,Enqvist:1997si,Enqvist:1998en}. However, this scenario requires additional investigation and lattice numerical simulations. Regardless of the order of the phase transition, domain walls resulting from the breaking of discrete symmetries can facilitate the trapping of fermions to form solitons~\cite{Macpherson:1994wf}. Moreover, they can even segregate baryons and antibaryons, offering a simultaneous explanation for both dark matter and the matter-antimatter asymmetry~\cite{Oaknin:2003uv,Atreya:2014sca,Liang:2016tqc,Ge:2017idw,Ge:2019voa,Zhitnitsky:2021iwg}.

\subsection*{Solitosynthesis}

The formation of solitons is sometimes called solitogenesis. After formation, the solitons can absorb free particles in the plasma during the cosmological evolution, a process known as solitosynthesis~\cite{Frieman:1988ut,Griest:1989bq,Frieman:1989bx,Bai:2022kxq}. In the case of Fermi-balls, solitosynthesis is active if $M/T$ is not excessively large, allowing abundant free $\chi$ particles in the plasma after the formation of solitons. The time of a Fermi-ball to accrete from a charge of $Q_1$ to $Q_2$ is estimated as~\cite{Bai:2022kxq}
\be\label{accretion_time}
\Delta t=\sum_{Q}\frac{1}{n_{\rm tv}\cdot\ave{\sigma v_{\rm rel}}}\sim\int_{Q_1}^{Q_2}\frac{\d Q}{n_{\rm tv}\cdot\pi R^2},
\ee
where $n_{\rm tv}$ is the number density of the free $\chi$ particles, and the absorption cross section $\ave{\sigma v_{\rm rel}}$ is approximately taken as the geometric limit $\pi R^2$. The maximum charge $Q_{\rm max}$ that a Fermi-ball can accumulate is estimated by the $Q_2$ gained within a few Hubble times, i.e. by setting $\Delta t\sim H^{-1}$. Utilizing the profile of the relativistic saturation limit in Table~\ref{tab:saturation} as a benchmark, the integration in \Eq{accretion_time} yields
\be
Q_2\lesssim Q_{\rm max}=\left\{Q_1^{1/3}+\frac16\sqrt{\frac{45}{\pi g_*}}\left[\frac{3}{16}\left(\frac{3}{2\pi}\right)^{2/3}\frac{1}{V(w_{\rm eff})}\right]^{1/2}\frac{M_{\rm Pl}}{T_*^2}n_{\rm tv}\right\}^3.
\ee
If the second term in the bracket is significant, Fermi-balls can experience substantial growth post-formation, but this evolution is highly dependent on the model. For instance, in cases involving FOPTs, a large $M/T_*\gtrsim\mO(10)$ is typical for particle trapping, and hence the free $\chi$ number density $n_{\rm tv}$ is suppressed by the Boltzmann factor $e^{-M/T_*}$. In this case, we often find that $Q_{\rm max}$ is nearly identical to $Q_1$, implying tiny change in Fermi-balls' charge throughout their evolution.

\subsection*{Evaporation}

In many models, Fermi-balls can dissipate heat by emitting light scalar quanta or through elastic scattering between constituent fermions and plasma particles, thereby tracking the Universe's temperature~\cite{Bai:2018dxf,Hong:2020est,Kawana:2022lba}. During the cosmic thermal history, the shape of $V (\phi,T)$ changes with temperature, affecting $Q_{\rm min}$, but the charge $Q$ of a given Fermi-ball remains fixed if solitosynthesis is inactive. Consequently, it's possible that $Q_{\rm min}(T) < Q$ at high temperature, leading to Fermi-ball formation; however, there comes a point at low temperature where $Q_{\rm min}(T) > Q$, causing Fermi-balls to evaporate and release free $\chi$ particles into the Universe.

Another type of evaporation is the decay of the constituent $\chi$ particles, possibly occurring primarily through the surface. This happens if the $U(1)_Q$ symmetry is slightly broken, allowing $\chi$ to be unstable. Such scenarios arise in models where $U(1)_Q$ is broken at high scales to generate a $\chi$-asymmetry, leading to $\chi$ decay at lower scales, although the lifetime can be very long. The decay process may involve the $\phi$ field, which complicates calculations involving background field theory, similar to the case of Q-ball surface evaporation~\cite{Cohen:1986ct}. If however the vertex is irrelevant to $\phi$, such as $y_\nu\bar\ell_L\tilde H\chi_R$ with $\ell_L$ and $H$ the SM left-handed lepton doublet and Higgs doublet respectively, then the calculation is simpler. In that case, since $m$ decreases deeper inside the Fermi-ball, it may be sufficient to approximate that only $\chi$ particles within a shell with a width of $m'^{-1}$ have the kinetic phase space to decay, where $m'$ is the mass of $\phi$ in the true vacuum of $V(\phi)$. Thus
\be
\frac{\d Q}{\d t}\sim-\frac{g_{\rm dof}}{6\pi^2}\frac{4\pi R^2\mu^3}{m'}\Gamma,
\ee
where $\Gamma$ is the single $\chi$ decay width calculating for $m=\mu$, and the number density in the shell is approximated as $g_{\rm dof}\mu^3/(6\pi^2)$. The lifetime is then $\tau\sim-Q\cdot\left(\d Q/\d t\right)$.
When the lifetime is comparable to the age of the Universe, Fermi-balls remain viable candidates for dark matter. The decay might generate the matter-antimatter asymmetry, analogous to some scenarios in the case of Q-ball~\cite{Enqvist:1997si,Enqvist:1998en}.

\subsection*{Collapse to primordial black holes}

It was proposed in Ref.~\cite{Kawana:2021tde} then extensively studied~\cite{Huang:2022him,Kawana:2022lba,Xie:2023cwi,Marfatia:2021hcp,Marfatia:2022jiz,Lu:2022jnp,Tseng:2022jta,Acuna:2023bkm,Chen:2023oew,Kim:2023ixo,Borah:2024lml} that the Fermi-balls may collapse to PBHs as they cool down. This argument is based on the energy profile, where $E\approx E_{\rm kin}+E_{\rm Yuk}$. The kinetic term $E_{\rm kin}\sim Q^{4/3}/R$ or $\sim Q^{5/3}/R^2$ is given by Table~\ref{tab:saturation} and the Yukawa term $E_{\rm Yuk}\approx-y^2Q^2f_y(1/m'R)/R$ is given by \Eq{EYUNR}, with $m'$ being interpreted as the effective mass of the scalar inside the soliton. As the temperature drops with the evolution of the Universe, $m'\approx\sqrt{m'^2_0+c\,T^2}$ decreases according to finite temperature field theory, leading to an increasing significance of the Yukawa energy because $f_y$ is monotonically increasing. When $E_{\rm Yuk}$ dominates the total energy, a Fermi-ball collapses into a PBH. However, the findings of this study dramatically alter this picture. The first key issue is the inconsistency of combining the formulae from the Yukawa attraction and saturation limits, as the former requires a small $Q$, while the latter is in the limit $Q\to\infty$, where the Yukawa term becomes negligible and cannot cause collapse.

One may wonder whether collapse can occur within the Yukawa attraction limit, where $Q$ deviates not far away from $Q_{\rm min}$, and the Yukawa interaction dominates. Addressing this, the second key issue emerges: it turns out that when $Q$ is small, Fermi-balls are not massive enough to collapse. This can be obtained by considering the ratio $\delta_{\rm Sch}=R_{\rm Sch}/R$ at the Coulomb limit, wherein the magnitude of Yukawa energy reaches its maximum. Here, $R_{\rm Sch}=2E/M_{\rm Pl}^2$ denotes the Schwarzschild radius, with $M_{\rm Pl}=1.22\times10^{19}$ GeV the Planck scale. $\delta_{\rm Sch}>1$ is necessary for collapse; however, \Eq{NRlimit} imposes an upper bound of
\be
\delta_{\rm Sch}<\frac{2Q}{R}\frac{M}{M_{\rm Pl}^2}=\left(\frac{4}{9\pi}\right)^{2/3}\left(\frac{g_{\rm dof}}{2}\right)^{2/3}\frac{y^2Q^{4/3}}{2\pi}\frac{M^2}{M_{\rm Pl}^2}<\frac{80\pi}{3}\left(\frac{w}{M_{\rm Pl}}\right)^2,
\ee
where the last inequality comes from requiring $E>0$ in \Eq{NRlimit}, which sets an upper limit on $Q$. Consequently, unless $w\sim M_{\rm Pl}$, collapse through Yukawa attractive force remains unattainable.

Although collapse driven by Yukawa binding force appears improbable, alternative pathways remain plausible. At the saturation limit, we observe $\delta_{\rm Sch}\sim Q^{2/3}$, suggesting collapse could occur if Fermi-balls accumulate adequate charge over their evolution~\cite{Lee:1986tr,DelGrosso:2024wmy}. In the FOPT-induced scenario, assuming the trapped fermions are relativistic, one obtains~\cite{Kawana:2021tde}
\be\label{Rratio}
\delta_{\rm Sch}\approx10^{-5}\times v_w^{2}\left(\frac{Y}{10^{-10}}\right)^{2/3}\left(\frac{H_*}{\beta}\right)^2\alpha^{1/2}.
\ee
Therefore, a large asymmetry with $Y\sim 10^{-3}$ and a slow and strong FOPT with $\alpha\sim\mO(10)$ and $\beta/H_*\sim\mO(1)$ might meet the Schwarzschild criterion and form PBHs. Furthermore, as the mass and radius of Fermi-balls in reality are extended functions~\cite{Lu:2022paj}, while \Eq{Rratio} solely estimates their mean values, there exists the possibility that the distribution tail of $\delta_{\rm Sch}$ surpasses 1, leading to the partial collapse of formed Fermi-balls into PBHs. The evolution of Fermi-ball profile due to changes in the shape of $V(\phi,T)$ could also lead to (partial) collapse. Another possible scenario is Fermi-balls form and accumulate huge charges at high temperatures where the thermal pressure also plays an important role in their stability, and then they cool down via emitting $\phi$ quanta and collapse to PBHs~\cite{Flores:2020drq,Flores:2023zpf}. When $\delta_{\rm sch}\sim1$, gravitational effects become significant, necessitating the techniques outlined in Refs.~\cite{Lee:1986tr,DelGrosso:2023trq,DelGrosso:2023dmv,DelGrosso:2024wmy} (also see Refs.~\cite{Balkin:2022qer,Balkin:2023xtr} for the case of interacting axions and nucleons inside white dwarfs and neutron stars).

Note that our discussion here pertains to the collapse of preexisting Fermi-balls, i.e., solitons formed via any process (not limited to FOPTs), which may undergo collapse during their evolution. This should be distinguished from scenarios where fermions are trapped within false vacuum remnants during a FOPT, directly collapsing into PBHs without forming solitons~\cite{Baker:2021nyl,Baker:2021sno,Cline:2022xhx,Lewicki:2023mik,Gehrman:2023qjn}.

\subsection*{Experimental signals}

Fermi-balls exhibit diverse cosmological and particle physics implications. If they endure as stable or long-lived objects until the present era, such compact balls can be probed by gravitational lensing~\cite{Marfatia:2021twj,Bai:2018vik,Bai:2018dxf}. The interactions between constituent fermions and the SM particles give rise to various particle physics phenomena, such as soliton capture by neutron stars or white dwarfs, soliton self-collisions, accretion of ordinary matter, and decay of the constituent fermions, all leading to detectable signals like $\gamma$-rays~\cite{Holdom:1987ep,Bai:2018vik,Bai:2018dxf,Forbes:2008uf,Forbes:2009wg,Gorham:2012hy}. It can be inferred that the astrophysical signatures become notably richer if the constituent fermions are the SM quarks (the quark nugget scenario). However, even if both $\chi$ and $\phi$ are SM singlets, the Higgs portal coupling $\phi^2|H|^2$ can generate observable effects. For instance, collisions between Fermi-balls may cause the emission of $\phi$ bosons, which subsequently decay into SM light particles.

The production mechanism of Fermi-balls yields extra signals. For instance, if Fermi-balls originate from a FOPT in the early Universe, then the stochastic gravitational waves (GWs) generated by the phase transition can help to probe the scenario~\cite{Hong:2020est,Marfatia:2021twj,Bai:2018vik,Bai:2018dxf}. Besides, the particle interactions required by the FOPT dynamics or particle trapping process may be probed at collider experiments through the production of $\phi$ and/or $\chi\bar\chi$~\cite{Huang:2022him}. Additionally, the presence of free $\chi$ particles that have penetrated the bubble wall to the true vacuum during the FOPT, though rare, could partly contribute to the dark matter abundance and serve as signals in direct detection experiments~\cite{Hong:2020est,Bai:2018dxf} (also known as the filtered dark matter scenario~\cite{Baker:2019ndr,Chway:2019kft,Chao:2020adk}).

Finally, we turn to the direct detection of Fermi-balls. The event rate passing through the detector can be estimated using
\be
N_{\rm dd}=\frac{\rho}{E}vL^2\Delta t\approx6\times\left(\frac{\rho}{\rho_{\rm dm}}\right)\left(\frac{10^{-4}~{\rm g}}{E}\right)\left(\frac{L}{10~{\rm m}}\right)^2\left(\frac{\Delta t}{1~{\rm yr}}\right),
\ee
where $L$ is the detector size, $\Delta t$ is the operating time, $\rho$ and $E$ are respectively the Fermi-ball energy density and mass today, $v\approx10^{-3}$ is the local velocity of Fermi-balls, and $\rho_{\rm dm}\approx0.376~{\rm GeV}/{\rm cm}^3$ stands for the local dark matter energy density. Thus, for light Fermi-balls and large detectors, the anticipated event rate is considerable, suggesting the potential detectability of such solitons through direct detection experiments. In fact, current direct detection experiments like Xenon1T or BOREXINO have already placed constraints on scalar solitons (Q-balls) involving the SM Higgs field~\cite{Ponton:2019hux,Jiang:2024zrb}, indicating the feasibility of probing Fermi-balls as well. Given that detectability relies on the interactions between Fermi-balls and detector materials, it is highly model-dependent, needing further investigations.

\section{Conclusion}\label{sec:conclusion}

This study establishes a framework for evaluating fermionic soliton profiles under a general scalar potential $V(\phi)$, incorporating the influences between fermion condensate and scalar field. This leads to the accurate derivation of soliton profiles under non-polynomial and complicated potentials, such as the logarithm potential $\sim \phi^4\log(\phi/w)$ in classically conformal theories~\cite{Iso:2009ss,Iso:2009nw,Chun:2013soa} or the non-analytical thermal corrections $\sim J_{B,F}(\phi^2/T^2)$ in finite-temperature field theories~\cite{Dolan:1973qd,Quiros:1999jp}. Soliton formation condition is given by the multi-vacuum structure of the effective potential $V_{\rm eff}(\phi)$, rather than solely relying on the bare scalar potential $V(\phi)$. Previous work has noticed $\ave{\bar\chi\chi}$ can generate a local minimum in $V_{\rm eff}(\phi)$ under a polynomial potential $V(\phi)=\lambda(\phi^2-w^2)^2/16$, where the authors call this non-perturbative vacuum scalarization~\cite{DelGrosso:2024wmy}. However, the generality of this feature is first shown in the current research. This novel insight reveals conventional fermion bound states as a specific category of fermionic solitons.

While the analytical formulae for fermionic soliton profiles commonly used in phenomenological studies are reproduced in the saturation limit of the results in this work, we emphasize that not all realistic formation mechanisms satisfy the saturation condition $Q\gg Q_{\rm cri}$. When the charge is not sufficiently large, it becomes necessary to evaluate the full profile of the soliton. Therefore one should be careful in case of the solitons production at high scales such as the Grand Unified Theory (GUT) scale or the seesaw scale. Another typical example is the Yukawa-induced-collapse PBH mechanism discussed in the literature, which this study finds challenging to realize, partly due to the Yukawa energy vanishing under the saturation limit.

This work can be extended in two primary directions. The first one is the determination of soliton profiles in more complicated situations. When discussing the soliton profiles in the early Universe, we replace $V(\phi)$ with $V(\phi,T)$ to include the thermal effects on the scalar potential. While this could be a good approximation, it overlooks finite temperature effects arising from the $\ave{\chi\bar\chi}$ source inside the Fermi-ball, as most formulae in this study assume a fermion distribution at $T=0$. However, extending our method to finite temperatures poses challenges. Utilizing the low-temperature expansion of the Fermi integral to evaluate the right-hand side of \Eq{Yuk_eom_1}, we obtain a $T^2$-dependent term in \Eq{Veff}
\be
V_{\rm eff}(\phi)\to V_{\rm eff}(\phi)-g_{\rm dof}\frac{T^2}{12}\mu\sqrt{\mu^2-m^2},
\ee
whose derivative diverges at the soliton boundary when $|m|=\mu$, thereby breaking perturbativity. Hence, more careful treatments are needed. Additionally, the results presented here solely consider the interaction between fermions and the scalar field. If $\chi$ is charged under some gauge group, the associate vector gauge fields should be included, introducing new features. Gauged fermionic solitons have been studied in the literature, primarily under a $U(1)'$ gauge group and polynomial scalar potential~\cite{Lee:1988ag,Anagnostopoulos:2001dh,Levi:2001aw}. A more precise discussion on the general gauge groups and scalar potentials would be valuable.

The second extension direction pertains to phenomenology. We have summarized three main formation mechanisms: direct fusion of free fermions, trapping fermions by walls (either FOPT bubble walls or domain walls), and scalar field fragmentation. While significant attention has been devoted to the fermion-trapping mechanism, further research is needed to explore the fusion and scalar fragmentation scenarios. Regarding fermion-trapping scenarios, it's intriguing to study deviations from the saturation limit, particularly for high-scale formed solitons. After formation, the fermionic solitons may grow through solitosynthesis, may evaporate to free $\chi$ fermions or SM particles, may collapse to PBHs, etc. There is ample opportunity for future exploration into the cosmological implications and experimental signals.

\acknowledgments

We would like to thank Philip Lu, Sida Lu, and Xun-Jie Xu for the very useful and inspiring discussions. This work is supported by the National Science Foundation of China under Grant No. 12305108.

\appendix
\section{Estimating the profile of Fermi-balls from a FOPT}\label{app:FOPT}

Here we mainly follow the method introduced in Refs.~\cite{Hong:2020est,Kawana:2021tde} to calculate the Fermi-ball formation dynamics. The decay rate of the Universe from the false vacuum $w'$ to the true vacuum $w$ is~\cite{Linde:1981zj}
\be
\Gamma(T)\sim T^4e^{-S_3(T)/T},
\ee
where $S_3(T)$ is the $O(3)$-symmetric Euclidean action evaluated from $V(\phi,T)$. Bubbles start to nucleate at $T_n$ when that the decay probability at a Hubble volume and Hubble time reaches 1, i.e. $\Gamma(T_n)H^{-4}(T_n)\approx1$. The temperature $T_p$ at which the true vacuum bubbles form an infinite connected cluster is called percolation, which happens when the true vacuum volume fraction reaches $0.29$~\cite{rintoul1997precise}. If the FOPT is not ultra-supercooled, $T_p$ is close to $T_n$, and either of them can be taken as the FOPT temperature $T_*$. For a FOPT happens in the radiation era, the action can be semi-analytically estimated as~\cite{Huber:2007vva}
\be
\frac{S_3(T_*)}{T_*}\approx168-4\log\left(\frac{T_*}{{\rm GeV}}\right)-4\log\left(\frac{\beta}{H_*}\right)+3\log v_w-2\log\left(\frac{g_*}{100}\right),
\ee
where $g_*$ is the effective degree of freedom of the relativistic particles in the plasma, and the parameters $\alpha$, $\beta/H_*$, and $v_w$ are explained in the main text. When $M-M'=y(w-w')\gtrsim\mO(10)\times T_*$, most of the $\chi$ particles cannot penetrate into the true vacuum. For example, $M'=0$, $M=15\,T_*$ and $v_w=0.4$ yields a trapping rate of $99.9\%$~\cite{Hong:2020est,Huang:2022him}.

The FOPT is usually described as the nucleation and expansion of true vacuum bubbles in the false vacuum background. However, at its final stage, it is more suited to be described as the shrinking and disappearing of the false vacuum remnants (or say ``bubbles'', ``pockets'') in the true vacuum background. With fermions trapped inside, those remnants eventually stop shrinking and form Fermi-balls. The size distributions of those false vacuum remnants can be calculated by the projection method~\cite{Lu:2022paj}, leading to an extended mass function of the Fermi-balls. However, the average values can be estimated by the following logic.

During evolution, false vacuum remnants first split to smaller ones, and then shrink to individual solitons when they are small enough. The critical size marking the transition from fragmentation to contraction is determined by the requirement that such a remnant should shrink to a negligible size before another true vacuum bubble nucleates inside it. Therefore, its size $R_*$ is estimated by
\be
\Gamma(T_*)\left(\frac{4\pi}{3}R_*^3\right)\left(\frac{R_*}{v_w}\right)\sim1,\quad\Rightarrow\quad R_*\sim\left(\frac{3v_w}{4\pi\Gamma(T_*)}\right)^{1/4}.
\ee
We assume the Fermi-ball formation occurs at the inverse scenario of percolation, specifically at the point where the false vacuum remnants can still amalgamate into an infinite connected cluster. This phenomenon arises when the false vacuum occupies 0.29 of the Universe's volume. Therefore the number density of the solitons at formation is given by $n_{\rm fv}^*=0.29/(4\pi R_*^3/3)$, which can be used to derive the number density today~\cite{Hong:2020est}.

Following the above discussions, the net fermions trapped in a single remnant is
\be
\ave{Q}\approx\begin{dcases}~Y\times\left(\frac{4\pi}{3}R_*^3\right)\times s_*,&\text{preexisting $\chi$-asymmetry;}\\
~\sqrt{\frac{g_{\rm dof}}{2\pi^2}M'^2T_*K_2\left(\frac{M'}{T_*}\right)\times\left(\frac{4\pi}{3}R_*^3\right)},&\text{thermal fluctuation.}\end{dcases}
\ee
where $Y$ is the $\chi$-asymmetry defined in the main text, $K_2$ is the second kind of modified Bessel function, and $s_*=2\pi^2g_*T_*^3/45$ is the entropy density. In case of $M'\to0$, one can use the formulae in Table~\ref{tab:saturation} to derive the numerical results listed in Table~\ref{tab:fopt_fbs} by setting $g_*\approx100$ and $g_{\rm dof}\approx2$.

Before closing, it is worth mentioning two implicit conditions in the FOPT-induced Fermi-ball formation scenario. First, the bubble expansion velocity $v_w$ should not be too close to 1, otherwise the fermions in the wall frame are very energetic with a kinetic energy of $\mO(\gamma_w T_*)\gtrsim(M-M')$ where $\gamma_w=(1-v_w^2)^{-1/2}\gg1$ is the Lorentz factor, and hence cannot be trapped in the false vacuum~\cite{Baldes:2021vyz,Huang:2022vkf,Chun:2023ezg}. Second, the reheating effect after the FOPT should not be prominent. If the energy released from transition reheats the Universe to $T_{\rm rh}\gtrsim (M-M')$, the Fermi-balls will be melt to free particles. Usually $T_{\rm rh}$ is estimated by $\sim(1+\alpha)^{1/4}T_*$, thus the Fermi-ball formation requires a not super-large $\alpha$. But even if $\alpha\gg1$, a low $T_{\rm rh}$ can be achieved when the decay width of $\phi$ is very small and hence the reheating is slow~\cite{Hambye:2018qjv}.

\bibliographystyle{JHEP-2-2.bst}
\bibliography{references}

\end{document}